\documentstyle[epsf,aps,twocolumn]{revtex}

\begin{document}
\draft

\twocolumn[\hsize\textwidth\columnwidth\hsize\csname @twocolumnfalse\endcsname

\begin{flushright}
{IFIC/99-31}
\end{flushright}

\title{High-quality variational wave functions for small $^4$He clusters}
\author{R. Guardiola and M. Portesi}
\address{Departamento de F\'{\i}sica At\'omica y Nuclear,
Facultad de F\'{\i}sica, 46.100-Burjassot, Spain}
\author{J. Navarro}
\address{IFIC (Centro Mixto CSIC Universidad de Valencia), 
Facultad de F\'{\i}sica, 46.100-Burjassot, Spain}
\date{\today}
\maketitle

\begin{abstract}
We report a variational calculation of ground state energies and
radii of $^4$He$_N$ droplets ($3 \leq N \leq 40$), using the
atom-atom interaction HFD-B(HE). The trial wave function has a
simple structure, combining two- and three-body correlation
functions coming from a translationally invariant
configuration-interaction description, and Jastrow-type
short-range correlations. The calculated ground state energies
differ by around 2\% from the diffusion Monte Carlo results.
\end{abstract}


]

\narrowtext

The research on liquid helium clusters has attracted a great
interest both experimentally and
theoretically\cite{toennies1,whaley1}. This research allows for
the analysis of the evolution of various physical properties for
increasing size of the system, going from single atoms to the
bulk.  Helium clusters are expected to remain liquid under all
conditions of formation, offering thus the possibility to study
finite-size effects in the superfluid state
\cite{sindzingre}. Moreover, it has been suggested that Bose
condensation could be detected by means of helium atom-cluster
collisions\cite{campbell}.  The experimental research has faced
the difficulties of size detecting the clusters.  Recently,
molecular beam diffraction from a transmission gratting
\cite{schoellkopf1} has proven to be successful to detect even
the $^4$He dimer\cite{schoellkopf2}, giving further impetus to
the study of helium clusters.

As the atom-atom interaction is well-known and relatively
simple, the solution of the Schr\"odinger equation has been
obtained using several microscopic methods, mainly based on
Monte Carlo
techniques\cite{pandha1,rama,barnett,chin1,lewerenz2}.
Variational Monte Carlo (VMC) calculations, using a Jastrow-like
ansatz for the many-body wave function, are currently used for
systems dominated by strong short-range interactions. The VMC
wave functions are the input for diffusion Monte Carlo (DMC),
Green function Monte Carlo or path integral techniques which
provide essentially exact, within statistical errors, ground
state energies of $^4$He clusters at zero temperature.
Conversely, these calculations constitute a useful benchmark to
test other many-body methods.

In this work we present a new method of obtaining high-quality
variational wave functions to describe small bosonic clusters.
The basic idea is to write the trial wave function as the
product of three terms, each of them with well defined roles.
The first term is the familiar two-body Jastrow correlation
factor, which controls the strong atom-atom repulsion at very
short distances.  The second term is related to a
single-particle description of the cluster and is written as the
product of $N$ single particle wave functions referred to the
center-of-mass coordinate; it provides the required confinement
of the constituents and fixes basically the size of the droplet.
Finally, the third term corresponds to a special version of the
configuration interaction (CI) expansion describing two- and
three-particle excitations, its role being to incorporate fine
details to the wave function for medium and long ranges, as well
as some collective effects.  In summary, the trial wave
functions we shall consider to describe the ground state of
$^4$He$_N$ clusters is
\begin{eqnarray}
&&\Psi({\bf R}) = \prod_{i<j} g(r_{ij}) \Phi({\bf R}) \nonumber \\
&&\hspace{5mm} \left( 1 + \sum_{i<j} f_2(r_{ij})  + 
\sum_{i<j<k} f_3(r_{ij}, r_{ik}, r_{jk}) \right),
\label{function}
\end{eqnarray} 
where $\bf R$ represents all single-particle coordinates. The
two-particle function $g(r_{ij})$ is the two-body Jastrow
correlation.  We shall use the form
\begin{equation}
g(r) = {\rm exp}\left( -\frac{1}{2}
\left(\frac{b}{r}\right)^{\nu} \right) ,
\label{mcmillan}
\end{equation}
introduced by McMillan\cite{mcmillan} many years ago for the
description of the homogeneous liquid using a 12-6 Lennard-Jones
interaction, with the value $\nu=5$ as required by Kato's cusp
condition.  Many VMC calculations have shown that this form is
also appropriate when dealing with interactions of the Aziz
type.  For instance, in Ref.\cite{chin1} fixed values $b=3$\AA,
$\nu=5$ have been used for the interaction HFDHE2\cite{aziz1},
independently of the number of constituents in the cluster.  It
makes sense to keep the same values for all the clusters,
because this pair correlation should be determined only by the
two-body interaction potential. We have slightly modified these
values, employing $b=2.95$\AA and $\nu=5.2$ in our calculations
with the interaction HFD-B(HE)\cite{aziz2}. Note that,
contrarily to other studies, we are considering for the Jastrow
term only the dominant part at very short-ranges.

The function $\Phi({\bf R})$ is the product of $N$
single-particle wave functions $\phi({\bf r}_i-{\bf R}_{CM})$,
referred to the center-of-mass of the system. If $\phi$ is a
gaussian (in other words, the 1s state of a harmonic oscillator
potential), it may be written as
\begin{equation}
\Phi({\bf R}) = \prod_{i<j} {\rm exp}\bigg( -
\frac{\alpha^2}{2N} r^2_{ij} \bigg), 
\end{equation}
depending on a size parameter $\alpha$.  Because of its
Jastrow-like form we may absorb it into the definition of the
Jastrow correlation factor, and define
\begin{equation}
\Phi_{J}({\bf R}) = \prod_{i<j} 
{\rm exp}\left( -\frac{1}{2}
\left(\frac{b}{r_{ij}}\right)^{\nu} - 
\frac{\alpha^2}{2N} r^2_{ij}
\right)
\label{jastrow}
\end{equation}

Our contribution to the variational description of the clusters
is the term within parentheses in Eq.(\ref{function}), with the
pair $f_2$ and triplet $f_3$ correlations, which we have called
above  configuration interaction and will be abridged as CI2
and CI3 respectively. In previous works\cite{bishop1} we have
extensively applied the CI2 scheme to the description of light
nuclear systems, with  less singular interactions than the usual
atom-atom potentials. The method is based on a linearized
version of the coupled cluster method (CCM)\cite{CCM} at the
SUB(2) and SUB(3) truncation approximation levels, restricted to
translationally invariant excitations up to three
particles-three holes. The linearized version of the CCM has
been shown to be equivalent to a special configuration
interaction scheme\cite{bishop2}, much more efficient than the
usual CI methods when dealing with realistic interactions. With
respect to CCM, the CI scheme loses the cluster property and the
correct scaling with the number of particles, being
unappropriate for extended systems. However it is much simpler,
as far as non-linear terms of the CCM expansion are absent.

The present mixed J-CI scheme may be compared with the
correlated basis functions (CBF) method as developed by
Feenberg, Clark and Krotscheck \cite{feenberg} for extended
systems, in which Jastrow correlations are combined with
non-orthogonal perturbation theory.  The key concept behind both
CBF and J-CI methods is to apply a divide-and-conquer strategy
to the determination of the correlated wave function.  It is
known that Jastrow correlations are very well suited to treat
the effects of the (strongly repulsive) short-range part of the
interaction, whereas a CI approach can effectively deal with
medium- and long-range correlations.  In our approach, we
substitute the standard configuration-interaction component by
our conceptually equivalent but significantly more effective
translationally invariant CI scheme.  Note that we are combining
additive CI and multiplicative Jastrow correlations.

The CI correlation functions $f_2$ and $f_3$ could be determined
by minimizing the ground state energy, resulting in a system of
coupled integro-differential equations for $f_2$ and $f_3$. As a
practical alternative we have expanded these functions in a set
of gaussians:
\begin{eqnarray}
&& 1 + \sum_{i<j}
f_2(r_{ij}) + \sum_{i<j<k} f_3(r_{ij}, r_{ik}, r_{jk}) =
\nonumber \\
&& \hspace{5mm} \sum_{p \leq q \leq r = 1}^{N_{\beta}} \, C_{p,q,r} \, {\cal S} \,
\left\{ \sum_{i<j<k}  {\rm e}^{-\beta_p r^2_{ij}}  
{\rm e}^{-\beta_q r^2_{ik}} {\rm e}^{-\beta_r r^2_{jk}} \right\}, 
\end{eqnarray}
where ${\cal S}$ indicates symmetrization with respect to the
particle labels. The gaussian expansion has proven to be a very
accurate representation of the correlation functions when both
negative and positive values of $\beta$ are included, with the
only restriction of having a square integrable wave function.
Among the set of  parameters $\{\beta_p \}$ we fix one of them,
say $\beta_1$, to zero. In this way we cover the three
possibilities contained in the ansatz wave function: restricting
the three labels $p,q,r$ to 1, we deal (up to a normalization
constant) with the correlated state $\Phi_J$; keeping only two
of these labels equal to 1, amounts to put on top of that state
the linear two-body correlations; finally the unrestricted
choice corresponds to the complete ansatz. Calculations
presented later on will be referred as J, J-CI2 and J-CI3
results, respectively.

A convenient short-hand notation for the gaussian expansion is
the following:
\begin{equation}
\sum_{\mu} C_{\mu} F_{\mu}({\bf R})
\end{equation}
where the subindex $\mu$ refers to the ordered set $(p \leq q
\leq r)$, and
\begin{equation}
F_{\mu}({\bf R}) = {\cal S}\, \left\{ \, \sum_{i<j<k} \exp
\left( -\beta_p 
r^2_{ij} -\beta_q r^2_{ik} -\beta_r r^2_{jk} \right) \, \right\}
\end{equation}
The variational determination of the energy reduces to the
solution of a generalized eigenvalue problem, which can be
stated as follows:
\begin{equation}
\sum_{\mu_2} \left( {\cal K}_{\mu_1,\mu_2} +
{\cal V}_{\mu_1,\mu_2} \right) 
C_{\mu_2} 
= E \sum_{\mu_2} {\cal N}_{\mu_1,\mu_2} 
C_{\mu_2} 
\label{eigenvalue}
\end{equation}
The matrix elements of the norm and the potential energy are the
integrals
\begin{equation}
{\cal N}_{\mu_1,\mu_2} = \int {\rm d}{\bf R}
|\Phi_J({\bf R}) |^2 F^*_{\mu_1}({\bf R}) F_{\mu_2}({\bf R}) 
\label{norma}
\end{equation}
and
\begin{equation}
{\cal V}_{\mu_1,\mu_2} = \int {\rm d}{\bf R}
|\Phi_J({\bf R}) |^2 F^*_{\mu_1}({\bf R})  \sum_{m<n}V(r_{mn})
F_{\mu_2}({\bf R})  
\label{poten}
\end{equation}
where $V(r)$ is the two-body interaction potential. For the
matrix elements of the kinetic energy operator we choose to
write them as
\begin{eqnarray}
&& {\cal K}_{\mu_1,\mu_2} = \int {\rm d}{\bf R}
|\Phi_J({\bf R}) |^2 F^*_{\mu_1}({\bf R}) \nonumber \\
&& \hspace{1cm} \frac{1}{\Phi_J({\bf R})} 
\bigg(-\frac{\hbar^2}{2m}\sum_n
\Delta_n \bigg) F_{\mu_2}({\bf R}) \Phi_J({\bf R}) 
\label{kine} 
\end{eqnarray}
since we shall use the positive definite function $|\Phi_J({\bf
R}) |^2$ as the guide of a Metropolis random
walk\cite{ceperley}.  Note that no substraction of the
center-of-mass contribution is necessary, as we are using a
translationally invariant wave function.  The number of unkown
amplitudes $C_{\mu}$ is $N_{\beta}+2 \atopwithdelims() 3$. As
mentioned above, the J and J-CI2 cases can be extracted easily
from the general wave function: indeed, they correspond to the
first $d \times d$ block of the matrices and the first $d$
amplitudes, with $d=1$ and $d=N_{\beta}$, respectively.

The last point to be discussed is the selection of the set of
parameters $\{\beta_p\}$. Our previous experience with nuclear
systems indicates that this selection is not very critical, as
far as a sufficiently large interval of length ranges $(1 /
\sqrt \beta_p)$ is included. In the calculations presented below
we have used the set $\{\beta_p/\alpha^2\} =
\{0,-0.05,0.5,1,4\}$. Large values of $\beta_p$ should not be
used, to avoid competition with the Jastrow factor.

The computational algorithm consists in carrying out a random
walk guided by $|\Phi_J({\bf R}) |^2$ as the probability
distribution function to evaluate the norm, potential and
kinetic matrices as given in eqs.(\ref{norma}-\ref{kine}). The
only adjustable parameter to be determined by minimization is
$\alpha$, since the amplitudes $C_{\mu}$ are self-adjustable.
A final warning is in order. It is well known that calculations
involving the solution of a generalized eigenvalue problem like
eq.(\ref{eigenvalue}) are prone to numerical instabilities,
because the over-complete basis giving rise to the overlap
matrix ${\cal N}$ may have a determinant close to zero. In our
case, the Monte Carlo evaluation of the overlap matrix
(\ref{norma}) results always in a positive matrix, and the
possible numerical instabilities are related only to numerical
rounding errors.

In Table I are collected our results for binding energies and
unit radii, defined as $r_0= \sqrt{5/3 \langle r^2 \rangle} /
N^{1/3}$, where $\langle r^2 \rangle$ is the square mean radius.
Columns labelled J, J-CI2 and J-CI3 correspond to the sequence
of trial wave functions previously explained.  Columns labelled
VMC and DMC display the results obtained by other authors using
the same interaction.

Our lowest order approximation (J) should be compared with
column VMC. It can be seen that our crude choice for the
two-body Jastrow correlation (\ref{mcmillan}) produces less binding than
the quoted VMC results.  The reason of these differences is
quite clear: once the parameters $b$ and $\nu$ have been fixed,
there remains a single variational parameter ($\alpha$) in
$\Phi_J$.  In contrast, the referred VMC calculations use more
elaborate forms for the trial wave function, including three-body
correlation functions in some cases, and containing from 5 to 10
parameters. Note however, that these differences decrease when
$N$ increases.

The inclusion of the CI correlation function $f_2$ changes
substantially the results.  It can be seen that figures in
columns VMC and J-CI2 are basically the same, within statistical
errors, except for the cluster with $N=40$.  The J-CI2 values
have been obtained by minimizing the energy with respect to
$\alpha$ and the set of five amplitudes describing $f_2$. The
resulting values of $\alpha$ are given in Table I. In some cases
we have also performed calculations including up to nine
gaussians. It turns out that the binding energy does not change
significantly, in spite of a considerably increase of the
computing time.

We finally refer to our best variational results, namely J-CI3.
There are no adjustable parameters in J-CI3, because we took the
value of $\alpha$ which minimizes the J-CI2 energy and the same
set of five $\beta$ exponents, the $C_{\mu}$ amplitudes being
determined by solving the generalized eigenvalue problem. It can
be seen in Table I that the ground state energies provided by
the J-CI3 scheme are above DMC ones by about 2\%, except at the
extremes of the table. For $N=3$ both J-CI3 and DMC energies
agree within statistical errors, and for $N=20$ and 40 the
differences are at the level of 4\%.  In any case, the J-CI3
description always shows a large improvement with respect to the
more elaborate VMC calculations displayed in Table I. Moreover,
the values obtained for the unit radius
are very close to the values given by DMC calculations.

It is interesting to show the sensitivity  to the value of the
size parameter $\alpha$ of the three levels of approximation.
As a typical example, in Figure  1 are shown our results for the
cluster with $N=8$.  For completeness this figure includes also
two horizontal dotted lines, which represent the optimal VMC and
DMC energies, and which are not related in any form to our
parameter $\alpha$. One observes that even if there is a strong
dependence of J energies with $\alpha$, this dependence almost
disappears when  $f_2$ and $f_3$ correlations are included.
The flatness of J-CI2 and J-CI3 energies is an indication of the
completeness of the basis used to describe these correlations.

In conclusion, we have presented in this paper a variational
wave function for $^4$He clusters. A simple Jastrow function of
McMillan type is sufficient to screen the strong short-range
atom-atom repulsion, whereas the remaining medium- and
long-range correlations seem to be adequately described by
linear CI correlation functions.  These functions are
authomatically determined once given the size parameter
$\alpha$, which is the only adjustable parameter.  We have shown
that these variational wave functions produce high-quality
results for small clusters. The method is likely applicable to
other bosonic systems.

\acknowledgements 
We are grateful to S.A. Chin and E. Krotscheck for providing us
with useful information about their previous work.  M.P.
acknowledges CONICET (Argentina) for a fellowship.  This work
has been partially supported by grant PB97-1139 (Spain).

\vspace{5cm}

\begin{figure}
\centerline{\epsfysize=2.in \epsfbox{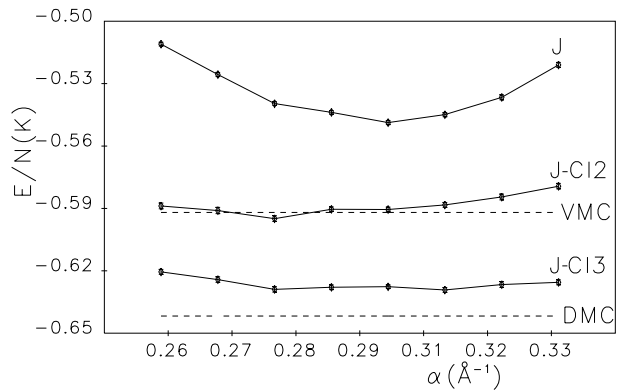}  }
\vspace{1pc}
\caption{Ground state energy per particle of cluster $^4$He$_8$,
as a function of the inverse length parameter $\alpha$, for the
sequence of trial wave functions J, J-CI2 and J-CI3. VMC and DMC
results are taken from Ref.\protect{\cite{lewerenz2}}.}
\label{Fig1}
\end{figure}

\newpage

\widetext

\begin{table*}
\caption{Ground state energies per particle  and unit radii  of
$^4$He$_N$ droplets, using the interaction HFD-B(HE). VMC and
DMC energies are taken from \protect{\cite{lewerenz2}} for
$N\leq10$, and \protect{\cite{whaley1}} otherwise. DMC radii are
taken from \protect{\cite{barnett}}.}
\begin{tabular}{c|ccc|cc|c|c|c}
  & \multicolumn{4}{c}{E/N (K)} &&& \multicolumn{2}{c}{$r_0$
(\AA)}  \\
\tableline
N & J & J-CI2 & J-CI3 & VMC & DMC & $\alpha$(\AA$^{-1}$)
& J-CI3 & DMC \\ 
\hline
3 & +.0136(4) & --.0373(8) & --.0430(10) & --.0385(3) & --.0436(2) &
.18 & 5.4(2) & 5.59 \\
4 & --.0835(7) & --.1326(13) & --.1398(15) & --.1333(10)& --.1443(2) &
.22 & 4.11(11) & 4.13 \\
5 & --.1984(8) & --.2502(14) & --.2616(13) & --.2506(3) & --.2670(3) &
.25 & 3.66(7) & 3.65 \\
6 & --.3187(11) & --.3678(12) & --.3868(11) & --.3676(2) & --.3950(2) & 
.27 & 3.42(6) &   \\
7 & --.4360(9) & --.4823(13) & --.5081(12) & --.4845(4) & --.5206(4) &
.27 & 3.31(5) & 3.22 \\
8 & --.5489(11)& --.5949(14) & --.6289(13) & --.5919(5) & --.6417(4) &
.28 & 3.18(4) &  \\
9 & --.6531(10) & --.6995(13) & --.7392(12) & --.6924(5) & --.7563(6) &
.28 & 3.11(3) & \\
10& --.7522(11)& --.7945(16) & --.8484(19) & --.7916(7) & --.8654(7) & 
.30 & 3.01(3) & \\
14&--1.0908(12)& --1.130(2)& --1.215(2)& --1.1290(7) & --1.2478(12)& 
.28 & 2.91(2) & 2.83 \\
20& --1.4743(16)& --1.5106(2)& --1.6336(15)& --1.510(2) & --1.688(2)&
.30 & 2.727(14)& 2.69 \\
40& --2.218(3)& --2.273(3)& --2.4563(14)& --2.430(2) & --2.575(3)&
.28 & 2.578(8) &   \\
\end{tabular}
\end{table*}

\end{document}